\documentclass[11pt,twoside]{article}

\usepackage{galev06}
\usepackage{epsf}
\usepackage{psfig}
\usepackage{lscape}

\begin{document}
\title{X-ray absorbed QSOs and the QSO evolutionary sequence}   
\author{M.J. Page}
\affil{UCL, Mullard Space Science Laboratory, Holmbury St. Mary, Dorking,
  Surrey RH5 6NT, UK}
\author{F.J. Carrera, J. Ebrero}
\affil{Instituto de F\'isica de Cantabria, Avenida de los Castros, 39005
  Santander, Spain}
\author{J.A. Stevens}
\affil{Centre for Astrophysics Research, University of Hertfordshire, College
  Lane, Hatfield AL10 9AB, UK}
\author{R.J. Ivison}
\affil{UK Astronomy Technology Centre, Royal Observatory, Blackford Hill,
  Edinburgh, EH9 3HJ, UK}

\begin{abstract}
Unexpected in the AGN unified scheme, there exists a population of
broad-line z$\sim$2 QSOs which have heavily absorbed X-ray
spectra. These objects constitute $\sim$10\% of the population at
luminosities and redshifts characteristic of the main producers of QSO
luminosity in the Universe. Our follow up observations in the submm
show that these QSOs are often embedded in ultraluminous starburst
galaxies, unlike most QSOs at the same redshifts and
luminosities.  The radically different star formation properties
between the absorbed and unabsorbed QSOs implies that the X-ray
absorption is unrelated to the torus invoked in AGN unification
schemes.  Instead, these results suggest that the objects represent a
transitional phase in an evolutionary sequence relating the growth of 
massive black holes to the formation of galaxies. 
The most
puzzling question about these objects has always been the nature of
the X-ray absorber.  We present our study of the X-ray absorbers based
on deep (50-100ks) XMM-Newton spectroscopy. We show that the
absorption is most likely due to a dense ionised wind driven by the
QSO. This wind could be the mechanism by which the QSO terminates the
star formation in the host galaxy, and ends the supply of accretion
material, to produce the present day black hole/spheroid mass ratio.
\end{abstract}



\section{Introduction}

The prevalence of black holes in present day galaxy bulges, and the
proportionality between black hole and spheroid mass \citep{merritt01} implies
that the formation of the two components are intimately linked. One way to
probe star formation in distant QSOs is to observe them at submm
wavelengths, and so measure the amount of radiation from young stars which is
absorbed and re-emitted by dust. With this in mind, we
observed 
matched samples of X-ray absorbed and unabsorbed QSOs at 850$\mu$m with
SCUBA.  These observations revealed a remarkable dichotomy in the submm
properties of these two groups of sources: almost all of the X-ray absorbed QSOs
at $z>1.5$  are
ultraluminous infrared galaxies, while the X-ray unabsorbed QSOs are not. This
suggests that the two types are linked by an evolutionary sequence, whereby the
QSO emerges at the end of the main star-forming phase of a massive galaxy
\citep{page04,stevens05,carrera06,stevens06}. 

However, the nature of the X-ray absorption remains
puzzling. 
These objects are characterised by hard,
absorbed X-ray spectra, and assuming that this is due to 
photoelectric absorption from cold material with solar
abundances, their column densities are $\sim 10^{22}$~cm$^{-2}$.
On the other hand, they have optical/UV spectra which are typical for
QSOs, with broad emission lines and blue continua. For a Galactic gas/dust 
ratio, the restframe UV spectra
would be heavily attenuated by such large columns of material, so the 
absorber appears to contain very little dust.

The X-ray absorption could be due to gas located within the AGN
structure, or from more distant material in the host galaxy, but in
either case the lack of dust is suprising. If the absorber were
associated with the obscuring dusty torus invoked in unification
schemes \citep{antonucci93} we would expect significant dust
attenuation, while the detection of the dust continuum at long
wavelengths implies that the interstellar media of the host galaxies
also have a high dust content. 
Therefore, in order 
to investigate the X-ray absorption, we have obtained deep (50--100ks) {\em
  XMM-Newton} observations of three submm bright, X-ray absorbed QSOs 
from our sample of hard-spectrum {\em
Rosat} sources \citep{page01a}. 

\begin{figure}
\begin{center}
\leavevmode
\psfig{figure=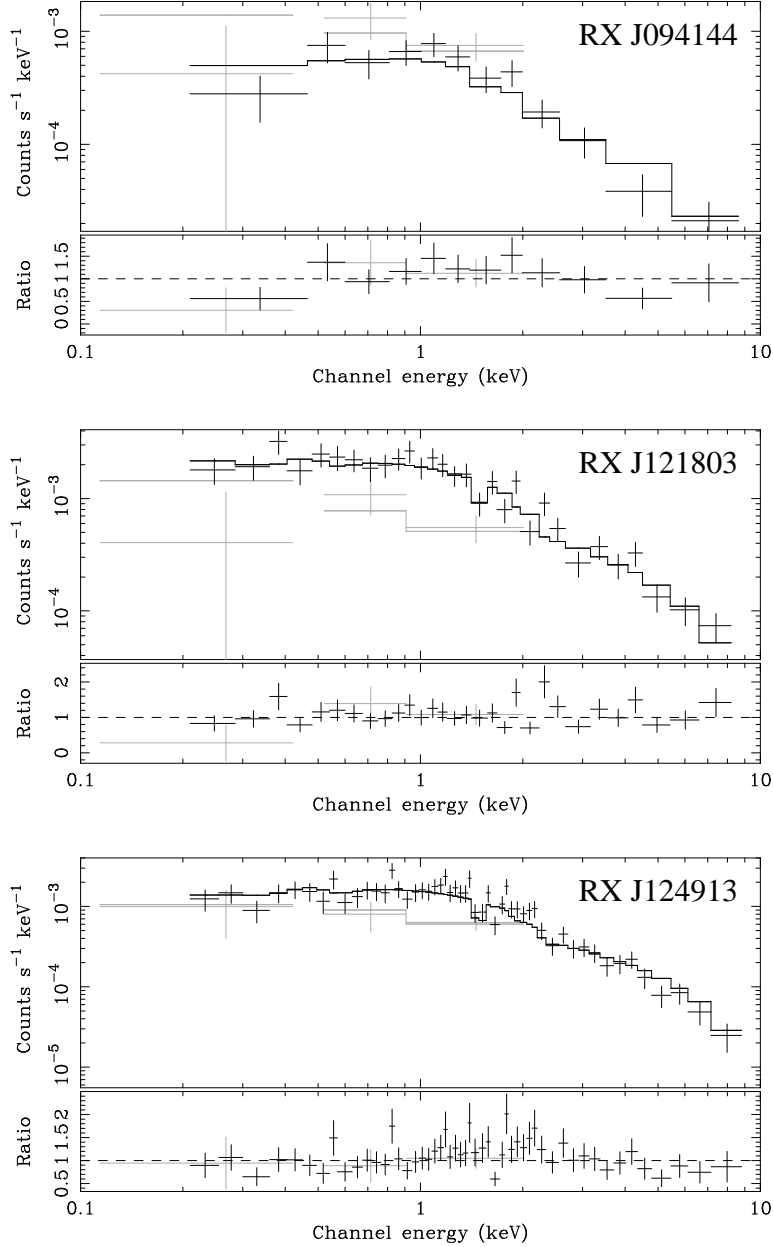,width=102truemm}
\caption{
{\em XMM-Newton} EPIC spectra (black) and {\em Rosat}
PSPC spectra (grey) of three X-ray absorbed QSOs.
The model is a simple power law with fixed Galactic absorption. 
The best-fit power law photon indices $\Gamma$ are $1.3\pm0.1$, $1.4\pm0.1$ and
$1.4\pm0.1$ for RX\,J094144, RX\,J121803, and RX\,J124913
respectively. Such photon indices are unusually hard for radio-quiet AGN. In 
all three objects there is a deficit of counts at the lowest
energy, indicating that absorption is responsible for the hard spectral shape;
this is also seen in the {\em Rosat} data. 
Furthermore, RX\,J094144 and RX\,J124913 show some systematic 
curvature relative to the power law model.
}
\label{fig:threeqsos}
 \end{center}
\end{figure}

\section{Results}

As a starting point, we fitted the {\em XMM-Newton} spectra with a power law 
and fixed
Galactic absorption. The best fit photon indices are unusually hard for 
QSOs: the QSOs have $\Gamma\sim 1.4$ (see Fig. \ref{fig:threeqsos}) 
compared to the 
$\Gamma=1.9$ that is typical for X-ray selected QSOs 
\citep[e.g.][]{mateos05,page06}. 
Furthermore, although the fits have acceptable $\chi^{2}/\nu$, the data
show a systematic deficit of counts relative to the model at 
the softest energies,
indicating that absorption is present. The original {\em Rosat} PSPC spectra
and the {\em XMM-Newton} spectra show excellent agreement 
(see Fig. \ref{fig:threeqsos}).

The hypothesis of a normal ($\Gamma=2$) AGN X-ray spectrum and a cold absorber
is strongly rejected for RXJ121803 and RXJ124913. Therefore we considered
ionised absorber models for the X-ray absorption, using the `xabs' model in
{\sc SPEX}, which includes both photoelectric and line absorption. For all
three AGN an acceptable fit can be obtained with a $\Gamma=2$ power law and an
absorber with an ionisation parameter log~$\xi \sim 2$ and column densities of
$10^{22.5}$--$10^{23.5}$~cm$^{-2}$. These absorbers have similar properties to
the high-ionisation absorber phases seen as outflows in some nearby Seyfert 1 
galaxies and
QSOs such as NGC\,3783 and PG\,1114+445 \citep{ashton04}.

At these ionisation parameters and column densities, the absorbers are likely
to originate in the AGN themselves, rather than in the host galaxies. The
detection of blue-shifted absorption lines in the rest-frame UV spectra of
several of our X-ray absorbed QSOs \citep{mittaz01} provides further
evidence that the ionised absorbers originate in outflows from the QSO. This
solution is attractive, because it is compatible with the lack of optical
extinction in these objects: if the absorber is driven as a wind, either from
the accretion disc or from evaporation of the inner edge of the molecular
torus, then dust will be sublimated before (or as) it enters the
flow. Identifying the X-ray absorption with an ionised wind also reconciles the
properties of the X-ray absorbed QSOs with geometric unification models,
because the submm detection statistics imply that the X-ray absorption
in these objects is not due to their orientation with respect to the dusty 
torus. 
Instead, the detection 
of an ionised wind in absorption implies that these objects, like optical or
X-ray selected QSOs in
general, are observed
pole-on rather than edge-on with respect to the torus.

\section{Implications for AGN and galaxy evolution}

As discussed by \citet{page04} and \citet{stevens05}, the low space density 
of X-ray absorbed QSOs relative to unabsorbed QSOs and to
distant ultraluminous galaxies detected in blank field SCUBA surveys, implies
that X-ray absorbed QSOs are caught during a short-lived transitional
phase. Before this brief phase, the AGN must be weak and heavily obscured, as
found by 
\citet{alexander05}; after this phase the host galaxy is essentially fully
formed, and the naked QSO shines brightly until its fuel is consumed. 
The origin of the X-ray absorption, and the brief duration of the transitional
phase were open questions from these studies. In \citet{page04} and 
\citet{stevens05} we suggested
that the transition is mediated by a wind from the QSO, which terminates 
the star formation in the host galaxy by driving gas and dust into the
intergalactic medium. Such a scenario
has also been proposed on theoretical grounds 
\citep[e.g.][]{fabian99,dimatteo05}. The EPIC
spectra of our X-ray absorbed QSOs suggest that the absorbers are
ionised winds driven by the AGN, and therefore that the transition between
buried AGN and naked QSO could indeed be mediated by a radiatively driven 
wind from the AGN.

%



\end{document}